\documentstyle[12pt,a4]{article}
\begin{document}
\begin{titlepage}
\begin{center}
\vskip .5cm

\hfill CERN--TH/95--70\\
\hfill hep-th/9503142\\

\vskip 1cm

{\Large{\bf A new class of \\
spatially homogeneous 4D string backgrounds}}

\vskip .5cm

{\bf  Nikolaos A. Batakis}
\footnote{Permanent address: Department of Physics, University of
Ioannina,
GR--45100
 Ioannina, Greece}
\footnote{e--mail address: batakis@surya20.cern.ch}
%\vskip .3cm

{\em Theory Division, CERN\\
     CH--1211 Geneva 23, Switzerland}

\vskip 3cm

{\bf Abstract }
\end{center}
\begin{quotation}\noindent
\end{quotation}
\newcommand{\ri}[1]{$#1(d\rightarrow)$}

A new class of spatially homogeneous 4D string backgrounds,
the \ri{X}  according to a recent classification,
is presented and shown to contain only five generic types.
In contrast to the case of $X(d\uparrow)$ (which
contains as a subclass all possible FRW backgrounds),
exact $SO(3)$ isotropy is always broken in the \ri{X} class. This
is due to the $H$-field, whose dual is necessarily along a principal
direction of anisotropy. Nevertheless, FRW symmetry can be attained
asymptotically for Bianchi-types $I$ and $VII_0$
in a rather appealing physical context.
Other aspects of the solutions found for types
$X=I,II,III,VI_{-1}$, and of the $VII_0$ case are briefly discussed.
\end{titlepage}

\newpage
\section{Introduction}

One of the major objectives of string theory in recent years
has been the development of a well-defined framework which
would upgrade the conventional
approach to cosmology near the Planck or string scale.
This is precisely the region where most of the major
cosmological problems arise and thus where increased insight
is needed most. The cosmology offered by string theory ought
to provide a sufficient understanding of that era, as well as
a subsequent `graceful exit' towards the
conventional (general-relativistic)
description of the more recent epochs. Such appears to be the
general motivation for the study of 4D string backgrounds, whether
they descend from a conformal field theory and higher-dimentional
compactifications, or simply satisfy the lowest-order string
beta function equations \cite{1}-\cite{4a}.
Next to the Friedmann-Robertson-Walker type of models,
the simplest backgrounds to be examined in the
above context are apparently those
which do not have SO(3) isotropy \cite{5} from the begining but attain that
state asymptotically during the later epochs of their
evolution. These belong to the wider category of spatially
homogeneous (but not necessarily isotropic) 4D string backgrounds
\cite{6}-\cite{7}, hereafter recaled as HSBs.

All possible HSBs have been recently classified
in a total of 576 cases \cite{8}, with a
major class $X(d)$ therein consisting of
those with `diagonal' metrics.
To justify the quotation marks we note that these metrics are always diagonal
only in certain {\em non-holonomic} frames, notably the one
supplied by the $\sigma^i$ forms which
respect homogeneity, namely they are
invariant under the left action of a {\em
transitive\/} $G_3$ group of isometries.
The group structure constants $C^i_{jk}$ which typically define
the commutation relations of its generators (here the Killing
vectors) also define these invariant 1-forms by $d\sigma^i=\frac{1}{2}
C^i_{jk}\sigma^j\wedge \sigma^k$.
The action of the group is {\em simply} transitive on its orbits
which are precisely the 3D hypersurfaces of
spatial homogeneity $\Sigma^3$ \cite{4},\cite{7}. This major class
$X(d)$ is actually subdivided into three classes,
the $X(d\rightarrow)$,
$X(d\nearrow)$ and $X(d\uparrow)$. The last one includes
the subclass $X(3d\uparrow)$
(here the number 3 shows that there is an extra $SO(3)$
{\em isotropy} group) which is of special interest because
it consists of all possible FRW bacgrounds.
The arrows just employed indicate
the orientation of the dual $H^\ast$ of the totally
antisymmetric field strength $H_{\lambda\mu\nu}$
with respect to the (pictured as horizontal) hypersurfaces $\Sigma^3$.
As we will see,
the $X(d\uparrow)$ class involves
$H^\ast$ congruences which are
{\em orthogonal} to $\Sigma^3$, that being
the only orientation
which has been examined in the literature on HSBs up to now.
In the present paper we study the $X(d\rightarrow)$ class,
namely HSBs with `diagonal' metrics and $H^\ast$
congruences which lie entirely within $\Sigma^3$.
We will also see that the presence of such a congruence forbids
the attainment of exact SO(3) isotropy so that any FRW-like behavior
would in principle be possible
in this class only at the limit of a vanishing $H$. Such is the physically
relevant case of an {\em asymptotically} vanishing $H$.

Our main results are presented in section 3, preceded
by some general definitions and preliminary findings in the
following section, and they
are further discussed in section 4.

\section{Preliminaries}

We want to examine spatially homogeneous 4D metrics which are
diagonal in the invariant $\{dt,\sigma^i\}$ basis, namely of the form
\begin{equation}
ds^2=-dt^2+a_1^2(t)(\sigma^1)^2+a_2^2(t)(\sigma^2)^2+
a_3^2(t)(\sigma^3)^2.\label{met}
\end{equation}
Such metrics will be considered here
as part of a solution for the background fields which satisfies
at least the lowest-order string beta-function equations
for conformal invariance. The scale factors (`radii\,') $a_i(t)$
along the principal directions of anisotropy are functions of the
time t only. Explicit holonomic expressions
for all bases $\{\sigma^i\}$ employed will be supplied
in the next section. Refering to the literature for details, we will
only add here that, depending on the structure
constants $C^i_{jk}$, two possible $G_3$-classes are
distinguished as $\cal A\;$ or $\cal B\;$ corresponding to whether
the adjoint representation of $G_3$ is traceless or not \cite{5}.
To further fix notation and conventions used,
we recall that the background field equations
can be derived from the effective action \cite{1},\cite{2}
\begin{equation}
S_{eff}=\int d^4x
\sqrt{-g}e^{\phi}(R-\frac{1}{12}
H_{\mu\nu\rho}H^{\mu\nu\rho}+\partial_{\mu}\phi\partial^{\mu}\phi-\Lambda)
\label{a}.
\end{equation}
In the  so-defined `sigma [conformal] frame' these equations are
\begin{eqnarray}
R_{\mu\nu}-\frac{1}{4}H_{\mu\nu}^2-\nabla_\mu\nabla_\nu\phi&=&0,
\label{b1}\\
\nabla^2(e^\phi H_{\mu\nu\lambda})&=&0, \label{b2} \\
\nabla^2\phi+(\nabla\phi)^2-\frac{1}{6}H^2+\Lambda&=&0,\label{b3}
\end{eqnarray}
where $\Lambda$ is the cosmological constant
emerging as a result of a non-vanishing
central charge deficit in the original
theory, hereafter set equal to zero. The fundamental fields
varied are the gravitational
$g_{\mu\nu}$,
the dilaton scalar $\phi$ and,
involved in the contractions $H_{\mu\nu}^2=H_{\mu\kappa\lambda}
{H_{\nu}}^{\kappa\lambda}
\, , H^2=H_{\mu\nu\lambda}H^{\mu\nu\lambda}$, the totally antisymmetric
field strenght $H_{\mu\nu\lambda}$. This field strength
(which may be equivalently
viewed here as a closed 3-form H) is defined in terms of the potential
$B_{\mu\nu}$ (equivalently the 2-form B) as
\begin{eqnarray}
H_{\mu\nu\rho}&=&\partial_\mu
B_{\nu\rho}+\partial\rho B_{\mu\nu}+\partial_{\nu} B_{\rho\mu} \label{H} \\
(H&=&dB) \nonumber
\end{eqnarray}
In addition to specific {\em coordinate} bases which we will introduce
later on as mentioned, we will also employ the general
orthonormal frame $\{\omega^\mu\}$ defined
so that (\ref{met}) is equivalently expressed as
\begin{equation}
ds^2=\eta_{\mu\nu}\omega^\mu\omega^\nu, \label{metd}
\end{equation}
with $\eta_{\mu\nu}$ the Minkowski signature $(-1,1,1,1)$.
We now observe that  $H_{\mu\nu}^2$
in (\ref{b1}) is diagonal iff $H^{\ast}$ has at most
{\em one} non-vanishing component in the $\{\omega^\mu\}$ frame.
Thus, for as long as one remains in the major class of `diagonal'
metrics (\ref{met}), the most obvious choice
involves a $H^\ast$ which is
orthogonal to the hypersurfaces of homogeity, namely
\begin{equation}
H^{\ast}=H^{\ast}_0\omega^0. \label{0}
\end{equation}
It it this choice which characterizes the $X(d\uparrow)$ class
and, as mentioned, the only one
which has been investigated in the literature on HSBs
so far. It is also the only choice which
would respect any existing or eventual $SO(3)$ isotropy in $\Sigma^3$.
Oviously however, other orientations of $H^\ast$
are also possible, in particular
along the complementary (transverse)
directions, as realized
\newcommand{\ri}[1]{$#1(d\rightarrow)$}
in the \ri{X} class which we examine in this paper. An apparently
fundamental feature in the latter case is the fact
that the allowed orientations of $H^\ast$ in $\Sigma^3$ are
severely restricted. A general restriction is that $H^\ast$
must be aligned with certain principal directions of
anisotropy. To precisely find which one(s), if any, we must examine
every possible isometry group (namely Bianchi type) separately.
The result of this examination may be summarized as follows.
In the simplest case of a fully abelian $G_3$
which is realized in Bianchi-type $I$ metrics,
$H^\ast$ is allowed to be along {\em any} one of the
three principal directions. This degeneracy is partly lifted
in the next case of type-$II$ metrics. For
any other Bianchi type there is either exactly one possible
orientation for $H^\ast$ or none at all.
As it turns out, for all metrics with isometry groups of
$G_3$-class $\cal A$
we are obliged to (or without loss of generality may) have
\begin{equation}
(G_3\,class\;{\cal A}):\,\,H^{\ast}=H^{\ast}_3\omega^3=
Aa_3^{-1}e^{-\phi}\omega^3 \label{3}
\end{equation}
where $A$ is a constant. For all other
cases there exist only two non-trivial possibilities, each one
unique for the indicated type, namely
\begin{eqnarray}
(type\,\,III):\,\,H^{\ast}&=&
H^{\ast}_2\omega^2=Aa_2^{-1}e^{-\phi}\omega^2 \label{2} \\
(type\,\,VI_{-1}):\,\,H^{\ast}&=&
H^{\ast}_1\omega^1=Aa_1^{-1}e^{-\phi}\omega^1 \label{1}
\end{eqnarray}
which we will adopt for the respective $G_3$-class $\cal B$ metrics.
Turning now to the dilaton field, one realizes that (\ref{b3})
can be significantly simplified and expressed as
\begin{equation}
\phi^{\prime \prime} =(H^\ast)^2\label{d}
\end{equation}
with a {\em prime} for $d/d\tau$.
The coordinate time $\tau$ has been defined by
\begin{equation}
dt = a^3e^{\phi}d\tau \label{tau}
\end{equation}
where the `volume' scale factor
\begin{equation}
a^3=a_1a_2a_3 \label{vol}
\end{equation}
determines the expansion of any co-moving volume element.
The universal time t can be explicitly given in
terms of $\tau$
once the $a_i(\tau),\phi(\tau)$ functions are known.
The set (\ref{b1}) for the metric coefficients in (\ref{met})
may now be expressed as
\begin{equation}
(\ln a_i^2e^{\phi})^{\prime \prime}+
(2V_i-(H^{\ast})^ 2\delta_{ij})a^6e^{2\phi} =0
\label{g}
\end{equation}
where $(H^{\ast})^ 2$ is the length of the dual
chosen from (\ref{3})-(\ref{1}). Each particular choice
is identified by one of the indices $j=3,2,1$, which also specify
(in that order) the particular principal
direction associated with it.
Each one of the Bianchi-type
depended potentials $V_i$ is in general a function of all three $a_i$.
The set of equations (\ref{g}) is subject to the initial value equation
\begin{equation}
(\ln a_1^2e^{\phi})^\prime (\ln a_2^2e^{\phi})^\prime +
(\ln a_2^2e^{\phi})^\prime (\ln a_3^2e^{\phi})^\prime +
(\ln a_3^2e^{\phi})^\prime (\ln a_1^2e^{\phi})^\prime +
(2\sum V_i-(H^{\ast})^ 2)a^6e^{2\phi}=
\phi^{\prime 2}, \label{gi}
\end{equation}
typically imposing a restriction on
the constants of integration.
Further restrictions in the form of {\em constraint equations}
emerge in the case of $G_3$-class $\cal B$
metrics \cite{5},\cite{7}.

\section{The $X(d\rightarrow)$ class of 4D HSBs}
\vspace{1cm}

We want to find all possible $X(d\rightarrow)$ HSBs, where
X specifies the Bianchi type, namely the isometry groups $G_3$
\cite{5},\cite{8} acting on the respective manifold.
As we have already implied, solutions exist
only in five cases, namely for
Bianchi types $I,II,III,VI_{-1},VII_{0}$.
With the exception of type $III$, for which a different expression
is involved, the dilaton field
is given by
\begin{equation}
e^\phi = Q^2e^{2P\tau}cosh A(\tau-\tau_0)  \label{f1}
\end{equation}
with $\tau_0,P,Q$ constants. It should be noted however
that, for $A=0$,
(\ref{f1}) reduces to
\begin{equation}
e^\phi = Q^2e^{2A_0\tau}
\label{f10}
\end{equation}
with $A_0$ a new constant not necessarily equal to $P$.
We will now present general solutions for types $I,II,III,VI_{-1}$,
while the $VII_0$ will also be examined
but in that case only special solutions (e.g., with
higher symmetry) seem attainable in closed form.

\vspace{1cm}
\noindent
{\boldmath \ri{I}}:$\;$
The metric (\ref{met}) may also be expressed in a coordinate basis as
\begin{equation}
ds^2=-dt^2+a_1^2(t)(dx^1)^2+a_2^2(t)(dx^2)^2+
a_3^2(t)(dx^3)^2 \label{I}
\end{equation}
and all $V_i$ vanish in (\ref{g}).
{}From the dual $H^\ast$ given by (\ref{3})
one finds the 3-form
\begin{equation}
H=A(a_1a_2)^2\,\,(d\tau \wedge dx^1\wedge dx^2), \label{H1}
\end{equation}
for the $H$ field which satisfies (\ref{b2}) and it is obviously exact.
It follows that a potential in (\ref{H}) would be
\begin{equation}
B=A(a_1a_2)^2x^2\,\,(d\tau \wedge dx^1).\label{B1}
\end{equation}
It is reminded that the dilaton field is given by (\ref{f1})
while from the set (\ref{g}) one can determine the $a_i$.
Thus, the rest of the solution is
\begin{eqnarray}
a_1^2e^{\phi} &=& Q^2e^{2(P+M)\tau} \nonumber \\
a_2^2e^{\phi} &=& Q^2e^{2(P-M)\tau} \nonumber \\
a_3^2 &=& L^2e^{2(P+N)\tau} \label{g1s}
\end{eqnarray}
with $M,N$ constants subject to the restriction
\begin{equation}
16P^2-4M^2+8PN=A^2 \label{rr1}
\end{equation}
as required by (\ref{gi}) and with L a numerical constant which
could be assigned any value. The set (\ref{g1s}) represents
a Casner-like solution \cite{5} which, together with the results
(\ref{H1}),(\ref{B1}),(\ref{f1}) describes the \ri{I} HSB.
This background exhibits asymptotic flat $(k=0)$ FRW
behavior with
\begin{eqnarray}
 M &=& 0 \nonumber \\
N &=& -\frac{\sqrt{3}+1}{4}A \nonumber \\
P &=& \;\frac{\sqrt{3}-1}{4}A, \label{il1}
\end{eqnarray}
as discussed in the next section.

\vspace{1cm}
\noindent
{\boldmath \ri{II}}:$\;$
The metric (\ref{met}) may also be expressed as
\begin{equation}
ds^2=-dt^2+a_1^2(t)(dx^1-x^3dx^2)^2+a_2^2(t)(dx^2)^2+
a_3^2(t)(dx^3)^2 \label{II}
\end{equation}
and it is obviously non-diagonal in this coordinate basis.
The $V_i$ in (\ref{g}) are
\begin{equation}
V_1=-V_2=-V_3=\frac{1}{2}(\frac{a_1}{a_2a_3})^2. \label{V2}
\end{equation}
{}From the dual $H^\ast$ given by (\ref{3})
one finds the 3-form
\begin{equation}
H=A(a_1a_2)^2\,\,(d\tau \wedge dx^1\wedge dx^2)\label{H2}
\end{equation}
for the $H$ field which satisfies (\ref{b2}) and it is obviously exact.
A possible potential in (\ref{H}) would be
\begin{equation}
B=A(a_1a_2)^2x^2\,\,(d\tau \wedge dx^1). \label{B2}
\end{equation}
The dilaton field is still given by (\ref{f1})
while from the set (\ref{g}) one can now find the $a_i$.
Thus, the rest of the solution is
\begin{eqnarray}
a_1^2e^{\phi} &=& M(\cosh M\tau)^{-1} \nonumber \\
a_2^2e^{\phi} &=& \frac{Q^4}{M}e^{4P\tau}\cosh M\tau \nonumber \\
a_3^2 &=& \frac{L^4}{M}e^{2N\tau}\cosh M\tau \label{g2s}
\end{eqnarray}
where $L,M,N,P,Q$ are constants,
not entirely arbitrary in view of the restriction
\begin{equation}
4P^2-M^2+8PN=A^2 \label{rr2}
\end{equation}
imposed by the initial value equation (\ref{gi}).

\vspace{1cm}
\noindent
{\boldmath \ri{III}}:$\;$
The metric (\ref{met}) may also be expressed in a coordinate basis as
\begin{equation}
ds^2=-dt^2+a_1^2(t)(dx^1)^2+a_2^2(t)(dx^2)^2+
a_3^2(t)(e^{x^1}dx^3)^2\label{III}
\end{equation}
There are non-vanishing potentials $V_1,V_3$ in (\ref{g}), namely
\begin{equation}
V_1=V_3=-\frac{1}{a_1^2}. \label{V3}
\end{equation}
This case involves a class-$\cal B$ isometry group
so we also have  the constraint equation
\begin{equation}
(\ln \frac{a_3}
{a_1})^\prime=0
\end{equation}
which (again without loss of generality) can be integrated to
\begin{equation}
a_1=a_3.
\end{equation}
{}From the dual $H^\ast$ given by (\ref{2})
one finds the 3-form
\begin{equation}
H=A(a_1a_3)^2e^{x^1}\,\,(d\tau \wedge dx^1\wedge dx^3) \label{H3}
\end{equation}
for the $H$ field which satisfies (\ref{b2}) and it is obviously exact.
It follows that a potential in (\ref{H}) would be
\begin{equation}
B=A(a_1a_3)^2e^{x^1}\,\,(d\tau \wedge dx^3). \label{B3}
\end{equation}
One can now proceed with (\ref{g}),(\ref{d})
to easily determine that
\begin{equation}
a_2=Qe^{P\tau},
\end{equation}
where $Q,P$ are constants, while the rest of these equations
reduce to
\begin{equation}
(\ln a_1^2e^{\phi})^{\prime \prime}-2Q^2e^{2P\tau}\,a_1^2e^{2\phi}=0,
\label{g3s}
\end{equation}
coupled to
\begin{equation}
\phi^{\prime \prime}-A^2a_1^4=0
\label{g3ss}
\end{equation}
A solution to this system is
\begin{eqnarray}
a_1^2 &=& a_3^2 = \frac{\sqrt{2}M}{A}(\sinh 2M\tau)^{-1} \nonumber \\
e^{2\phi} &=& \frac{3MA}{\sqrt{2}Q^2}e^{-2P\tau}(\sinh 2M\tau)^{-1}
\end{eqnarray}
where $7M^2=P^2$ as required by the initial value equation (\ref{gi}).
The above solution was not obtained by quadratures, so we have no
proof that it is the most general to the (\ref{g3s}),(\ref{g3ss}) system.

\vspace{1cm}
\noindent
{\boldmath \ri{VI_{-1}}}:$\;$
The metric (\ref{met}) may also be expressed in a coordinate basis as
\begin{equation}
ds^2=-dt^2+a_1^2(t)(dx^1)^2+a_2^2(t)(e^{-x^1}dx^2)^2+
a_3^2(t)(e^{x^1}dx^3)^2. \label{VI}
\end{equation}
The only non-vanishing potential in (\ref{g}) is
\begin{equation}
V_1=-\frac{2}{a_1^2}. \label{V6}
\end{equation}
It should be noted, however, that in this case
we also have the constraint equation
\begin{equation}
(\ln \frac{a_3}
{a_2})^\prime=0
\end{equation}
which can be integrated
(essentially without loss of generality) to
\begin{equation}
a_2=a_3
\end{equation}
{}From the dual $H^\ast$ given by (\ref{1})
one finds the 3-form
\begin{equation}
H=-Aa_2^4e^{x^1}\,\,(d\tau \wedge dx^1\wedge dx^3), \label{H6}
\end{equation}
for the $H$ field which satisfies (\ref{b2}) and it is obviously exact,
so that a possible potential in (\ref{H}) would be
\begin{equation}
B=Aa_2^4x^2e^{x^1}\,\,(d\tau \wedge dx^3).\label{B6}
\end{equation}
The dilaton field is again given by (\ref{f1})
while the essentially remaining $a_1$ function can be obtained
by direct integration in
the set (\ref{g}) with (\ref{V6}). We thus find the rest
of the solution which may be expressed as
\begin{eqnarray}
& & a_1^2 = L^2\exp (M\tau+\frac{Q^2}{4P^2}e^{4P\tau})  \nonumber \\
& & a_2^2e^{\phi}=a_2^2e^{\phi} = Q^2e^{2P\tau},
\label{g6s}
\end{eqnarray}
where $L,M,Q,P$ are constants.
These are not entirely arbitrary in view of the restriction
\begin{equation}
8P^2+8PM=A^2 \label{rr6}
\end{equation}
imposed by the initial value equation (\ref{gi}).

\vspace{1cm}
\noindent
{\boldmath \ri{VII_0}}:$\;$
The metric (\ref{met}) may also be expressed in a coordinate basis as
\begin{equation}
ds^2=-dt^2+a_1^2(t)(dx^1)^2+a_2^2(t)(dx^2)^2+
a_3^2(t)(dx^3)^2 \label{VII}
\end{equation}
and the $V_i$ in (\ref{g}) are
\begin{equation}
V_1=-V_2=\frac{a_1^4-a_2^4}{2a^6},\;\;\;
V_3=-\frac{(a_1^2-a_2^2)^2}{2a^6}. \label{V7}
\end{equation}
{}From the dual $H^\ast$ given by (\ref{3})
one finds the 3-form
\begin{equation}
H=A(a_1a_2)^2\,\,(d\tau \wedge dx^2\wedge dx^3), \label{H7}
\end{equation}
for the $H$ field which satisfies (\ref{b2}) and it is obviously exact,
so that a possible potential in (\ref{H}) would be
\begin{equation}
B=A(a_1a_2)^2x^2\,\,(d\tau \wedge dx^3)\label{B7}
\end{equation}
It is reminded that the dilaton field is given by (\ref{f1})
and, as in the two other $G_3$-class $\cal A$ cases, we also have
\begin{equation}
a_1a_2e^\phi = Q^2e^{2P\tau}.
\end{equation}
However, it seems unlikely that the general \ri{VII_0} HSB can be found
in closed form from the remaining integration of (\ref{g}) with (\ref{V7}).
We observe, however, that under the $SO(2)$ partial isotropy realized if
$a_1=a_2$, these equations reduce to the $SO(2)$-symmetric type-I set,
which we have already at our disposal from (\ref{g1s}). In other words,
the backgrounds $VII_0(2d\rightarrow)$ and $I(2d\rightarrow)$
are identical, and so are their respective subcases
which involve asymptotic flat FRW behaviour. The same behavior
is expected from the general \ri{VII_0} bacground, under a
choice of constants analogous to (\ref{il1}). The above results
will be further discussed in the next section.

\section{Conclusions}

The \ri{X} class of 4D HSBs has been investigated, with explicit
solutions found for all but the last one of five possible
generic cases, realized at $X=I,II,VI_{-1},III,VII_0$. For the last
type only specialized (that is, with more symmetry)
solutions could be found in closed form, such as the
$VII_0(2d\rightarrow)$ which essentially coincides
with $I(2d\rightarrow)$. Although there are no FRW backgrounds
in \ri{X} (unless on goes over to trivial limits), there exists a
subclass therein with
{\em asymptotic} flat FRW behavior. As we will see shortly, this subclass
can be viewed as a counterpart of the FRW models. It can be
obtained from \ri{I} and \ri{VII_0} with proper choice of
constants, such as the one in (\ref{il1})
made for the type-I case. This choice
may appear as a {\em fine-tuning}, but such a characterization
is in a sense misleading. To appreciate this rather important
point, we may now compare with the case of
the FRW models. As recently shown, the far richer $X(d\uparrow)$ class
contains as a subclass all possible FRW models \cite{7}.
Every one of these models is
obtained by a specific choice of constants (entirely analogous to
the one made in (\ref{il1})) which is equivalent to introducing three
extra Killing isometries, namely
the $SO(3)$ isotropy of the FRW regime. It is
then clear that the choice (\ref{il1}) involves the same or
perhaps even {\em less} of a fine tuning because it introduces three
{\em asymptotic} Killing vectors \cite{5}. The impossibility
to establish full (rather than just asymptotic) $SO(3)$ isotropy
in \ri{X} is of course due to the presence of the $H$ fiefld,
visualized by its $H^\ast$ congruence within the hypersufaces
of homogeneity $\Sigma^3$. On physical grounds such behaviour
may be more interesting than the FRW one, in
view of the {\em dynamical} attainment of isotropy predicted
for the later epochs in the cosmic evolution.

Other gross physical aspects of the solutions presented here
firstly include a confirmation of the
expected presence of an initial singularity.
One can also establish the absence of inflation in the entire
\ri{X} class. It should be noted, however, that
the latter result has been established only in
terms of the scale factor $a$ introduced in (\ref{vol}),
because inflation has actually {\em not} been studied in the context
of anisotropic cosmology
(incidentally, this default is also a paradox
if one recalls that, for example, the horizon problem is intimately
related to the issue of spatial anisotropy) \cite{5}.
It should also be noted that any string background may appropriately
be viewed as the counterpart of a general relativistic {\em vacuum}.
In that sense, many \ri{X} configurations which by our present
findings are charactrized as non-existent
(namely those with $X=IV,V,VI_{h},VII_h,VIII,IX$), could be realized
in the presence of appropriate sources added to
the effective action (\ref{a}).

It is conceivable that some of the backgrounds presented here
may discend from a CFT and, further, one would
like to examine the behavior of these backgrounds
under various duality transformations (cf., eg., \cite{4a} and
refs cited therein). Here we will briefly comment
on the type of new backgrounds which can be obtained
under {\em abelian} target-space duality. In the $X(d\uparrow)$
class, duality transformations largerly reproduced
backgrounds with metrics in the same class \cite{7}. In the present
case the situation is complementary, in the sense that
the \ri{X} HSBs in
general poduce duals with metrics outside that class. At the same
time, the original symmetry can be
severely reduced or virtually lost \cite{4a}.
We also observe that due to the presence of
non-vanishing $B_{0i}$ components of
the potentials in (\ref{B1}),(\ref{B2}) etc.,
the new metrics will involve $g_{0i}$
components so that even if a $\Sigma^3$ has survived
as a hypersurface of homogeneity, it will cease to also
be a hypersurfaces of simultaneity. One consequence of that
would be the impossibility to define a cosmic time t in the
respective manifold. These aspects should be examined in
detail, especially in the context of ref. \cite{4}.

\end{document}